\documentclass[conference]{IEEEtran}
\IEEEoverridecommandlockouts
\usepackage{cite}
\usepackage{amsmath,amssymb,amsfonts}
\usepackage{algorithmic}
\usepackage{graphicx}
\usepackage{textcomp}
\usepackage{xcolor}
\usepackage{framed}

\usepackage[hidelinks]{hyperref}

\def\BibTeX{{\rm B\kern-.05em{\sc i\kern-.025em b}\kern-.08em
    T\kern-.1667em\lower.7ex\hbox{E}\kern-.125emX}}
\begin{document}

% Watching Vision Videos: Alone or Together
% Vision Video: Viable Viewing Options
% Distributed Stakeholders
%
\title{Viewing Vision Videos Online: \\Opportunities for 
Distributed Stakeholders\\
}

\author{\IEEEauthorblockN{Lukas Nagel}
\IEEEauthorblockA{\textit{Leibniz University Hannover} \\
\textit{Software Engineering Group}\\
Hannover, Germany \\
lukas.nagel@inf.uni-hannover.de}
\and
\IEEEauthorblockN{Jianwei Shi}
\IEEEauthorblockA{\textit{Leibniz University Hannover} \\
\textit{Software Engineering Group}\\
Hannover, Germany \\
jianwei.shi@inf.uni-hannover.de}
\and
\IEEEauthorblockN{Melanie Busch}
\IEEEauthorblockA{\textit{Leibniz University Hannover} \\
\textit{Software Engineering Group}\\
Hannover, Germany \\
melanie.busch@inf.uni-hannover.de}
%\and
%\and
%\and
}

\maketitle

%\begin{abstract}
%context&motivation -> question&Problem -> principle/ideas/results -> contribution (+ nettes Häppchen)
%Creating shared understanding between stakeholders is essential for the success of software projects. Misaligned mental models can lead to conflicting requirements, which in turn hinder the development process. The use of videos to present abstract visions of a system and its functionality is one approach to counteract this problem. These videos are usually shown in in-person meetings. However, stakeholders might be unable to attend such a conjoined meeting. Scheduling difficulties are worsened by necessary travel when stakeholders come from different locations. Methods for the use of vision videos in online settings are necessary. Furthermore, methods enabling an asynchronous use of vision videos are needed for cases when conjoined meetings are impossible even in an online setting.

%In this paper, we compare synchronous and asynchronous viewings of vision videos in online settings. The two methods are piloted in a preliminary experiment resulting in multiple advantages and disadvantages for each type. Our results point to relevant use cases for both methods. We also discuss how our findings can be applied to the elicitation of requirements from a crowd of stakeholders.
\begin{abstract}
Creating shared understanding between stakeholders is essential for the success of software projects. Conflicting requirements originating from misaligned mental models can hinder the development process. The use of videos to present abstract system visions is one approach to counteract this problem. These videos are usually shown in in-person meetings. However, face-to-face meetings are not suited to every situation and every stakeholder, for example due to scheduling constraints. Methods for the use of vision videos in online settings are necessary. Furthermore, methods enabling an asynchronous use of vision videos are needed for cases when conjoined meetings are impossible even in an online setting. 

In this paper, we compare synchronous and asynchronous viewings of vision videos in online settings. The two methods are piloted in a preliminary experiment. The results show a difference in the amount of arguments regarding the presented visions. On average, participants who took part in asynchronous meetings stated more arguments. Our results point to multiple advantages and disadvantages as well as use cases for each type. For example, a synchronous meeting could be chosen when all involved stakeholders can attend the appointment to discuss the vision and to quickly resolve ambiguities. An asynchronous meeting could be held if a joint meeting is not feasible due to time constraints. We also discuss how our findings can be applied to the elicitation of requirements from a crowd of stakeholders.

%However, the ongoing Covid-19 pandemic has made it impossible to conduct in-person meetings where vision videos would normally be discussed. As a consequence, misalignments between stakeholders' mental models and the vision presented in the video can not be resolved. Therefore, methods for the use of vision videos in times of social distancing and home office are necessary.
\end{abstract}


\section{Introduction}
%intro
The elicitation of requirements is one of the key phases in the requirements engineering process \cite{paetsch2003requirements}. In this phase, requirements from all relevant stakeholders must be considered. However, in situations where key aspects of the system to be developed are still unclear, conflicting requirements can be elicited from different stakeholders \cite{van2000requirements}. It is important to create a shared understanding of the desired system between stakeholders to solve such conflicts and to keep them from threatening the project's success. Videos present one possible medium of creating shared understanding. In these so-called vision videos, concrete visions of the functionality of a system are presented. Stakeholders watching the videos can detect, discuss and resolve misalignments of their mental models \cite{karras2020representing}.

%This means that misalignments of stakeholder's mental models can be detected, discussed and resolved. %It is paramount that all relevant stakeholders are included in this process to ensure a projects success REF.
Once vision videos are created, stakeholders usually attend a conjoined meeting in which a vision video is shown and discussed \cite{brill_videos_2010, Schneider2017}. However, stakeholders can not always attend such meetings in person. They can be distributed among different locations, time zones or even continents. In these cases, vision videos need to be watched and discussed online. Online meetings make it easier for requirements engineers to collaborate with distributed stakeholders since no travel arrangements have to be made. Nevertheless, finding a time that is suitable for every stakeholder can be a challenging or sometimes even impossible task. This problem only increases in importance with larger numbers of relevant stakeholders, for example when eliciting requirements from a crowd~\cite{groen2015towards, todoran2013cloud}. Discussing vision videos with only a subset of all relevant stakeholders introduces the risk of missing out on valuable ideas, especially those that build upon suggestions of others. 

%Objective
The overall goal of this research is to \textit{find a viable method to conduct vision video meetings in online settings}. Hence, we investigate synchronous and asynchronous viewings of vision videos in online settings. Synchronous viewings are the closest adaption of traditional vision video meetings. We look to find advantages and disadvantages of their use in an online context. Asynchronous viewings can be used when it is impossible to find a suitable time for a conjoined meeting. Therefore, we seek to explore the applicability of distributed viewing of vision videos to requirements engineering processes with crowds of stakeholders. The online setting allows many more people, or a crowd, to watch vision videos and provide feedback on the visions presented.

%Contribution
In this paper, we introduce concepts for synchronous and asynchronous viewings. The two methods are tested in a preliminary experiment. Our results indicate that vision videos can still lead to meaningful discussions regardless of the online setting. Furthermore, we find asynchronous viewings to facilitate the use of vision videos for crowds of stakeholders. The approach enables large numbers of stakeholders to generate ideas and express concerns about illustrated visions without inhibitions.

The rest of this paper is structured as follows: Section \ref{sec:rw} presents related work. We introduce our methodology including a preliminary experiment in section \ref{sec:methodology}. Results of the experiment are laid out in section \ref{sec:results} and discussed in section \ref{sec:disc}. The paper is concluded in section \ref{sec:concl}.

\section{Related Work}
\label{sec:rw}
%%% WRITING FLOW %%
%% Video and Vision Video in RE -> CrowdRE -> (Online Learning) -> Group vs. Individual Decision Making -> Karras Related Work -> Research Gap

Videos have been used in requirements engineering for many years. 
Creighton et al.~\cite{creighton_software_2006} proposed ``software cinema'', a scenario video creation methodology.
Such videos show workflows of customers which are not implemented yet. Xu et al. \cite{xu_user_2013} also examined scenario videos in the scope of software development and applied them in the context of a case study.
Brill et al.~\cite{brill_videos_2010} have listed opportunities for using videos in different phases of requirements engineering. 
They have conducted a study to investigate whether video creation is effective and efficient in comparison to use cases.

Gathering feedback on videos from a crowd in online settings has already been applied before the Covid-19 pandemic. 
Keimel et al.~\cite{keimel_qualitycrowd_2012} have proposed a web application to collect opinions from a crowd regarding video quality. 
For more complex feedback, Lasecki et al.~\cite{lasecki_glance_2014} developed a video coding tool with a conversation interaction paradigm between user and system. A user asks questions in natural language and receives answers from crowd workers. Crowd opinions are aggregated as final feedback.

Crowd opinions can be gathered from a group or from multiple sessions with individuals. In the field of psychology, several researchers studied the decision-making mechanisms of groups. 
Gruenfeld et al.~\cite{gruenfeld_group_1996} have tested whether group decision making is dependent on the degree of homogeneity within the group. 
Toma and Butera~\cite{toma_hidden_2009} provided selected information to influence a group's decision making. 
Clart et al.~\cite{clark_comparing_2015} have conducted a game to compare decision making of groups and individuals. Their results indicate that groups perform better than individuals with regard to decision rationality. 
In the engineering domain, Ezin et al.~\cite{ezin2019} have built a video recommendation system for individuals based on group preferences. 
All the literature listed indicates that group decision making is a complex process that seems to depend on group members, external information, and individual opinions.
%In our work, we wanted to know in which extend the decision making as a group and as an individual differs by collecting feedback from vision video.

During the pandemic, Karras et al.~\cite{karras2020using} used vision videos in a virtual focus group of five participants. 
%Within the scope of their work they elicited requirements from two users. As a first step, they used silent vision videos with live audio commentary from the moderator. After that, both users watched the video again separately without any explanation from the moderator. Lastly, parts of the videos are replayed in the two participants group for discussion. 
Based on their experiences, they formulated recommendations for the use of vision videos in virtual focus groups.

In our work, we use vision videos for the evaluation of different visions. We collect feedback from potential stakeholders on different system visions in online meetings. The stakeholders are asked to choose one of the presented visions and provide reasons for their decision. Our efforts focus on the viewing of vision videos online in group settings or individually. We examine whether differences between these settings can be observed.

%Decisions can be made from group discussion and individual interview. In the field of requirements engineering, we will investigate if and how decision making in a group and as an individual differ.

\section{Methodology}
\label{sec:methodology}
The goal of this research is to \textit{find a viable method to conduct vision video meetings in online settings}. For this purpose we designed two different methods, namely synchronous and asynchronous viewing.

%The goal of this research as defined in the goal definition template by Wohlin et al.~\cite{wohlin_experimentation_2012} is the following:
%Analyze different methods of watching vision videos for the purpose of finding the best suited method with respect to the feedback elicited from the viewpoint of stakeholders in the context of online settings.
%For this purpose we designed two different methods, namely synchronous and asynchronous viewing.

\subsection{Synchronous Viewing}
\label{sec:sync}
Synchronous viewing of vision videos in online settings aims directly at simulating the experience of traditional meetings. For this purpose, stakeholders join a video conferencing tool like Zoom\footnote{\url{https://zoom.us/meetings}} or BigBlueButton\footnote{\url{https://demo.bigbluebutton.org/gl}}. Here, they can use a microphone to speak to other attendees. Webcams can also be used to enable stakeholders to see one another. The mentioned video conferencing tools allow users to share videos within them. Vision videos can be shown and controlled by the meeting's organizers. The control elements change the video for all attendees simultaneously. This means that pauses, rewinds or jumps to specific parts of the video are visible for all participants. Meeting organizers can therefore control the focus of the discussion. Video sections that are unclear can be revisited and discussed explicitly by using these controls.

\subsection{Asynchronous Viewing}
\label{sec:async}
A different method of watching vision videos in online settings is to do so asynchronously. Asynchronous viewing of videos means that each stakeholder watches the video by themselves. There is no fixed time at which all stakeholders have to watch simultaneously. Instead, a deadline can be given. In this method, all control over the video lies with the stakeholder. Asynchronous viewing requires additional means to record stakeholder’s ideas or concerns.

\subsection{Goal and Research Questions}
\label{sec:study}

According to the goal definition template by Wohlin et al.~\cite{wohlin_experimentation_2012}, we formulated our research goal as follows:  
\begin{framed}\noindent
\textbf{Goal definition:} \newline \textbf{We analyze} different methods of viewing vision videos
\newline \textbf{for the purpose of} finding a viable method
\newline \textbf{with respect to} the feedback elicited
\newline \textbf{from the viewpoint of} stakeholders
\newline \textbf{in the context of} an online experiment setting.
\end{framed}

%RG: find the best way to conduct vision video meetings in online settings
%\subsection{Research Questions}
Based on our research goal, we formulated the following research questions: 
\begin{itemize}
    %\item \textbf{RQ1:} What is the best way to conduct vision video meetings in online settings?
    \item \textbf{RQ1:} What are the advantages and disadvantages of viewing vision videos synchronously?
    \item \textbf{RQ2:} What are the advantages and disadvantages of viewing vision videos asynchronously?
    %\item \textbf{RQ1:} What are the advantages and disadvantages of synchronous viewings of vision videos?
    %\item \textbf{RQ2:} What are the advantages and disadvantages of asynchronous viewings of vision videos?
    
    %\item \textbf{RQ2:} Can the individual choice of vision video variants be influenced by discussions with others?
    %how influential/impactful is the discussion with other stakeholders during synchronous meetings?
\end{itemize}

\begin{figure*}[!h]
\centering
\includegraphics[scale=0.58]{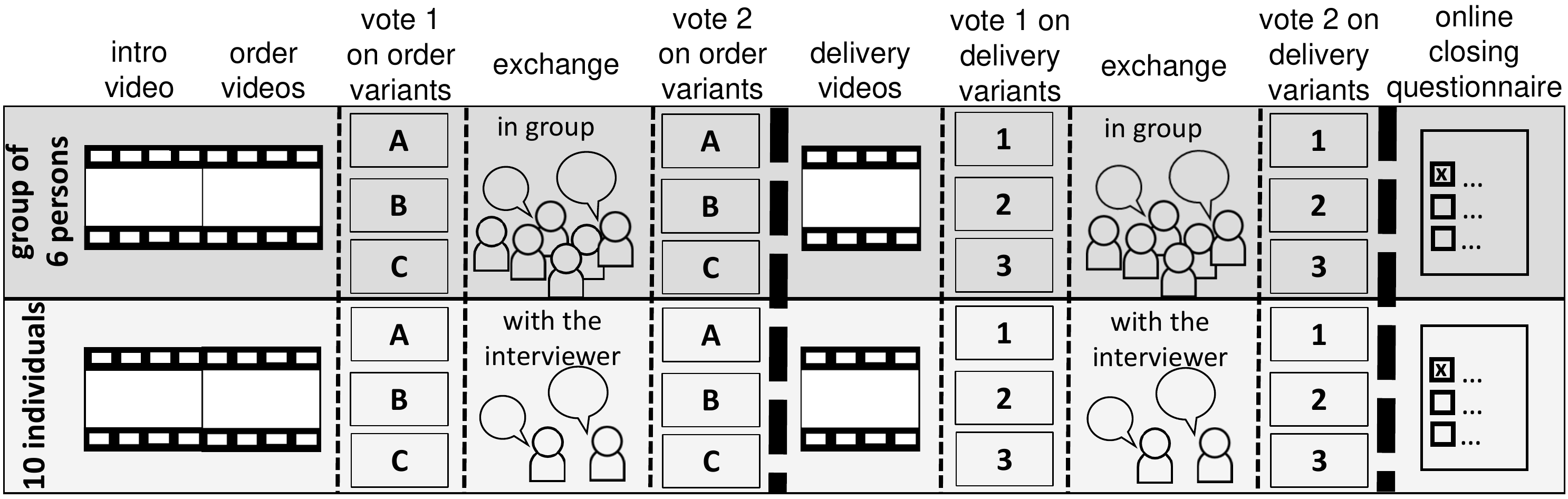}
\caption{Experiment design}
   \label{fig:Experiment Design}
\end{figure*}

\subsection{Hypotheses}
%Diskussionen bei synchronen Meetings unterstützen den RE Prozess in großem Maße H1_0
%Asynchrones Schauen führt zu mehr Feedback einzelner Teilnehmer H2_0

%Asynchrones Schauen führt zu mehr Argumenten pro Teilnehmer --> zu welcher Forschungsfrage?
%Diskussionen beim synchronen Schauen führen zu mehr Meinungsänderungen --> zu welcher Forschungsfrage?
Based on the two methods of synchronous and asynchronous viewing of vision videos, we conducted an experiment with two sets of participants: a group (synchronous viewing) and individuals (asynchronous viewing). 
Therefore, we analyzed the following hypotheses:
%\newline \textbf{H1\textsubscript{1}}: There is a difference between the groups regarding a change in the choice of variants after the discussion. 
\newline \textbf{H1\textsubscript{0}}: There is no difference between the sets of participants regarding a change in the choice of variants after the discussion.
%\newline \textbf{H2\textsubscript{1}}: There is a difference between the groups regarding the amount of feedback given by single participants. 
\newline \textbf{H2\textsubscript{0}}: There is no difference between the sets of participants regarding the amount of feedback given by single participants.

The respective alternative hypotheses {H1\textsubscript{1}} and {H2\textsubscript{1}} state the opposite of the null hypotheses, namely that there is a difference between the sets of participants.

Both hypotheses aim at possibly existing differences in the two online settings considered. Based on these differences we formulate advantages, disadvantages and potential use cases.

\subsection{Experiment Design}
\paragraph{Material}
In our experiment, we used a video that we divided into two thematic sections by pausing it. At the beginning of the first video section, there was an introduction to familiarize viewers with the topic of shopping in rural areas and the potential challenges associated with it. After this introduction, the first video section showed three possible options for ordering products: by taking a picture, by pressing a button, or by automatic measurement. The second section of the video includes three delivery options for products by receiving the package from a neighbour, by drone delivery, or by dropping the package in the trunk of a car. The videos used in our experiment were produced and already used in the context of the paper ``Refining Vision Videos'' by Schneider et al.\cite{schneider_refining_2019}.
At the end of the experiment, participants were asked to complete a LimeSurvey\footnote{\url{https://www.limesurvey.org/de/}} questionnaire. The questionnaire contained demographic questions and questions about the experiment. 

\paragraph{Participant Selection}
All 16 participants of our experiment took part voluntarily. The participants were between 21 and 33 years old (M = 25.8, SD = 4.1). Three of the participants were female and thirteen male. 
Ten of the participants are students; six are currently employed or have been employed in the past. Fourteen of the participants are studying subjects in STEM\footnote{STEM (Science Technology Engineering Mathematics)} fields or are currently working or have worked in STEM fields in the past.
\begin{figure}[!h]
\centering
\includegraphics[scale=0.46]{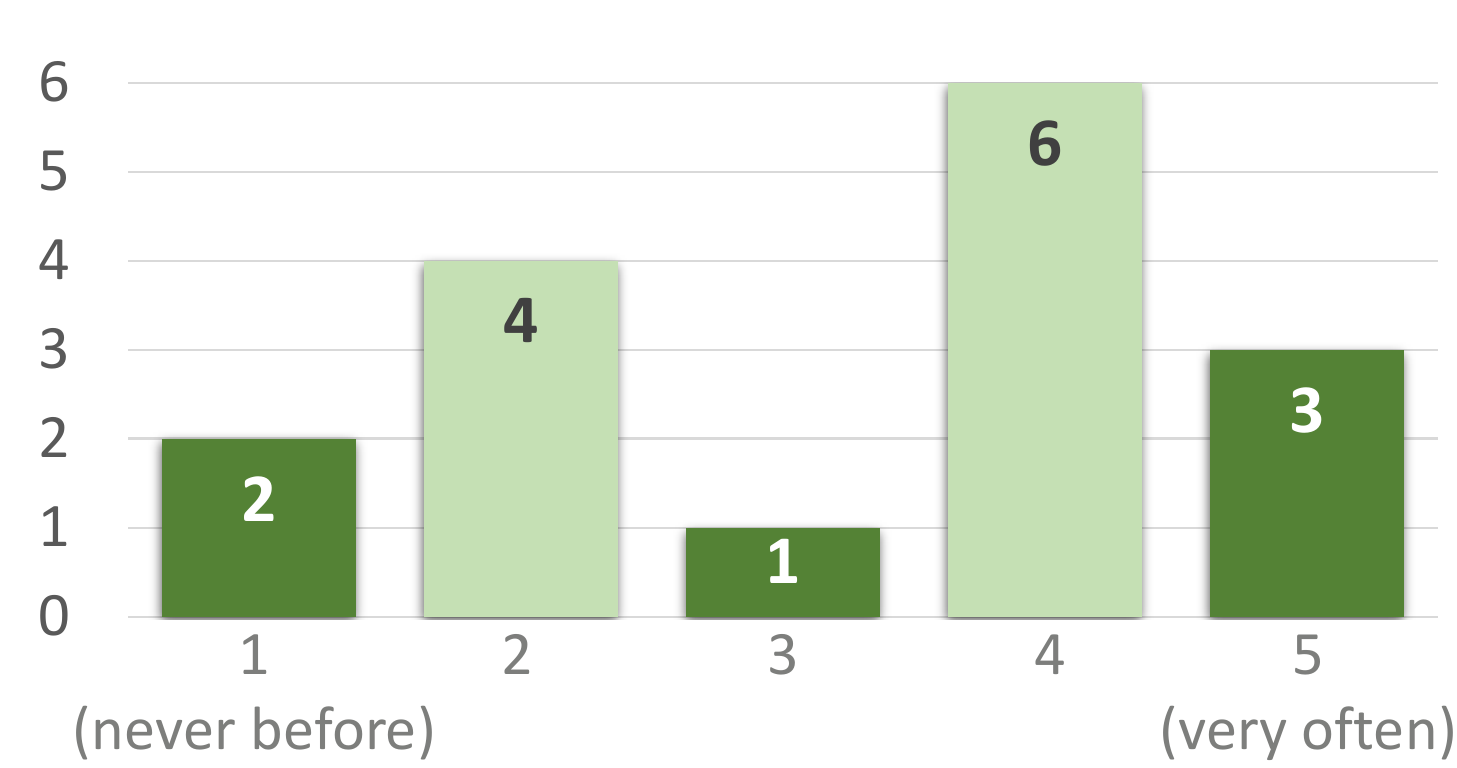}
\caption{How often have you been involved in software development processes?}
\label{fig: Demographics}
\end{figure}

Figure \ref{fig: Demographics} shows the distribution of responses to the question \textit{How often have you been involved in software development processes?}. On a likert scale labeled from 1 (never before) to 5 (very often), two participants stated that they had never been involved in a software development process before. %Four participants stated a two, one participant stated a three, six participants stated a four, and three participants stated a five. 

%\begin{figure}[!h]
%\centering
%\includegraphics[scale=0.57]{Figures/Participants_Demographics_1.pdf}
%\caption{How often have you been involved in software development processes?}
%\label{fig: Demographics}
%\end{figure}

\paragraph{Experiment Procedure}
At the beginning of the planning phase of the experiment participants signed up for appointments in an online calendar. Subsequently, a date was set for the group session on which at least six participants had registered. The other participants were assigned to the group of individual sessions.   
Prior to the experiment, a consent form and an overview were sent to the participants by e-mail. Six of the participants took part in a group session, the other ten participants took part individually. The experiment was conducted virtually in BigBlueButton. 
Our design is divided into two segments. Figure \ref{fig:Experiment Design} shows the sequence of sections for the individuals and the group of six participants. 

%\begin{figure*}[!h]
%\centering
%\includegraphics[scale=0.58]{Figures/StudyDesign4.pdf}
%\caption{Experiment design}
%   \label{fig:Experiment Design}
%\end{figure*}

At the beginning of our experiment the participants saw the thematic intro section and the three order variants (labeled A, B and C). After watching the videos, participants voted on which variant they would choose via a short poll started by the experimenter. 
Then the group participants discussed their choices among each other. The individuals were asked about pros and cons of each variant by the interviewer. After this exchange, participants voted again on which order variant they would choose via a short poll started by the experimenter. This sequence of video watching, voting and an exchange was repeated in the second segment regarding the delivery variant videos (labeled 1, 2 and 3). 
At the end of the experiment, the participants were asked to complete an online questionnaire.

\paragraph{Data Analysis Procedures}
In our experiment, we collected data in three different ways. We have documented the participants' votes of the variants, a total of four votes for each participant. In addition, we analyzed the audio recordings of the sessions to be able to count the mentioned pros and cons for each order and delivery variant. The last data source we used are the results of the LimeSurvey questionnaire. Regarding the results of the questionnaire, we cleaned the data by removing incomplete response data-sets.\\
\textit{Analyzing the Choice of Variants}: To examine our data regarding the first hypothesis, we analyzed the results of two questions of the online questionnaire in a descriptive way. Within the scope of these questions, we asked the participants whether they had changed their opinion regarding the choice of variants after the discussion and if so, why. \\
\textit{Analyzing the Amount of Arguments}: To answer our second research question, we listened to the audio recordings of the sessions to count the arguments for each variant. Based on the results of the counting, we calculated the average arguments per participant for both groups. \\
\textit{Analyzing Additional Results}: For the analysis of the additional results (\textit{Video Quality, Changes of Opinion, Chosen Variants}) the results of the online questionnaire were used. The respective average of the data, the median and the standard deviation were calculated for each subsection.

\section{Results}
\label{sec:results}
Analyzing the data as described leads to the following results:

\subsection{Choice of Variants}
We asked our participants to choose one of the shown variants four times. The exit survey included a question regarding whether or not participants had changed their choices. We also asked about the reasons for these changes. %The vision video used in the experiment consisted of two sets of variants. One set presented three possibilities (labeled A, B and C) for the design of an ordering process. The other showed three possibilities (labeled 1, 2 and 3) for the manner in which the product is delivered to a customer.
%The vision video shown offered two sets of three variants each, one for the manner in which a product is ordered and one for the manner in which it is delivered to a customer. 
%For both sets, we asked our participants to choose a variant immediately after seeing them in the video. After a discussion we asked them again. In this way, we measure whether or not their choice had changed. We also asked our participants whether or not their choice had changed based on the discussion in the exit survey. For members of the synchronous meeting (group session), this question referred to their exchange with other participants. 
%Individual participants watching the video asynchronously were told that this question concerned their exchange with the researchers. In these exchanges they were asked to express advantages and disadvantages of individual variants.
% Vorschlag: For members of asynchronous meeting, they were asked to make a decision for the second time after discussion with researcher. This discussion should influence the second decision, because participants expressed advantages and disadvantages of individual variants.  

Four out of six participants (66\%) of the group session indicated that their choices had changed. When asked for the reasons behind these changes, they mentioned misunderstandings being cleared up and critical questions being answered in the discussion with other attendees.

As for the asynchronous viewers partaking in the individual sessions, six out of ten test subjects (60\%) changed their opinion. Three of them added that the additional time they spent thinking about the advantages and disadvantages of individual variants was the reason for these changes. Another participant mentioned that they needed time to adjust to their role as a stakeholder for a system concerning rural infrastructure. Based on these results, we reject \textbf{H1$_0$} and accept \textbf{H1$_1$}.

\subsection{Arguments}
During the experiment we counted the arguments in favor and against individual variants mentioned by participants. %In the following, we list argument examples:  For the order variant A (by taking pictures on mobile), the two advantages ``minimum cost", ``flexibility" were stated by the participants. 
For example, the advantages ``minimum cost" and ``flexibility" were stated by participants regarding order variant A in which a picture of the product is taken using a smartphone. An example for a mentioned disadvantage of the delivery by drone was that ``It could be problematic if I were not at home when the drone comes".
%For the delivery variant 2 (by drone), one disadvantage was mentioned: ``It could be problematic if I were not there when the drone came".
The group attending the synchronous meeting gave a total of 24 arguments. With six participants taking part in the meeting, this means that each participant mentioned four arguments on average. Furthermore, we can calculate the average amount of arguments for each variant shown in the video per participant. With six variants present, this results in 0.66 arguments per variant and participant.

For the ten individual test subjects partaking in asynchronous viewings, we counted a total of 151 arguments. This means that an average of 15.1 arguments per participant was given. As the same video was shown, the average number of arguments per variant and participant can be calculated in the same manner resulting in a value of 2.5.

\begin{figure}[ht]
\centering
\includegraphics[scale=0.98]{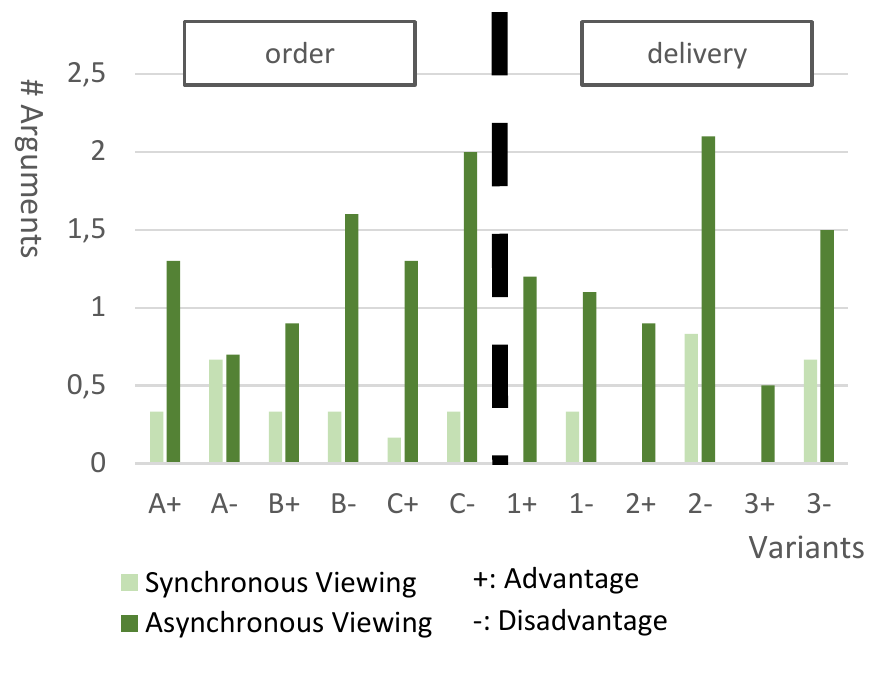}
\caption{Average number of arguments per variant and participant}
\label{fig: NumberOfArguments}
\end{figure}

Figure~\ref{fig: NumberOfArguments} shows that the average number of arguments per variant and participant from those taking part in the asynchronous viewings (dark green) is far higher than the corresponding data from the synchronous viewing (light green).  
%The increased number of arguments per variant per participant 
This difference represents an increased amount of feedback. Therefore, we can not accept \textbf{H2$_0$}. Instead, we accept \textbf{H2$_1$}.

Based on the given arguments, we have also collected new requirements from participants.
One participant suggested scanning a QR-Code as an alternative to ordering by taking a picture (variant A), because the product can be recognized precisely.
While criticizing the missing recipient on drone delivery (variant 2), one participant proposed that the landing time and place should be arranged before delivery. 
Apart from the requirements presented, numerous enhanced or new requirements were stated by the participants.
Based on the results of our preliminary experiment, we infer that listing the advantages and disadvantages can support the elicitation phase. 

%könnte hier noch ergänzen, dass die meisten Argumente Contra-Argumente waren, aber was unterstützt das in unserer Argumentation?

\subsection{Additional Results}
\label{addtional-result}
In addition to the results of the experiment, we asked participants further questions in an exit survey on their experiences during the experiment. We collect their answers to gather information about the validity of our results.

%\paragraph{Video Familiarity}
%One question of our survey asked whether or not our participants were familiar with the video used for the study. Not a single participant had seen the video or its variants before taking part in the experiment. %None of our participants were familiar with the video used for the study before participating in it.

\paragraph{Video Quality}
One question of the exit survey asked about participant's agreement with a statement expressing that the video's quality was good enough to understand its contents. With an answer of 1 indicating full disagreement and a rating of 5 indicating full agreement, participants gave an average rating of 4.13 (M = 4, SD = 0.96). Notably, 75\% of participants agreed with the statement. Only a single test subject disagreed.

\paragraph{Changes of Opinion}
The survey included a question regarding participant's self-reported tendencies to change their opinions after exchanges with others. With an answer of 1 meaning that they were very unlikely to change their opinion, while 5 indicated a high probability, participants' answers provided an average of 2.56 (M = 2.5, SD = 0.96).

\paragraph{Chosen Variants}
%TODO: sollte ich das auch noch auf die beiden Gruppen aufteilen?
Finally, we asked participants whether they were satisfied with the variants chosen in the experiment. Again, ratings of 1 and 5 indicated strong dissatisfaction and strong satisfaction respectively. The participants of our experiment answered with an average of 3.25 (M = 4, SD = 0.93). Notably, the strongest ratings of 1 and 5 were not chosen by any test subjects.

\section{Discussion}
\label{sec:disc}
%asynchrones viewing hat aber auch vorteile, vor allem mehr argumente und längere Bedenkzeiten
%Vorschlag: Videos sollten asynchron vorher gesehen und dann in Runde diskutiert werden -> Konflikte und weiterspinnen von Ideen

Based on observations made by researchers during the preliminary experiment and the collected results, we gathered insights about our research questions.
The results of our preliminary experiment reveal insights on our research questions.
%We conclude this paper by discussing the results of our preliminary study regarding our research questions. Threats to validity are also addressed.

\subsection{Pros and Cons of Synchronous Viewing (RQ1)}
We trialed synchronous viewings of vision videos in online settings as described with a group of six participants. Based on our observations and the results of the survey we found the following advantages and disadvantages:

%Synchronous Watching
\paragraph{Advantages}
Viewing vision videos in a synchronous manner resembles traditional meetings most closely. Processes that worked for vision video meetings in person can be adopted for synchronous online meetings with minimal adjustments.

Synchronous viewings of vision videos mean that stakeholders attend the same meeting. They can directly discuss the contents of the shown video. This means that misaligned mental models can quickly be identified and cleared up. The exchanges can also result in new ideas. Stakeholders can express their opinions and suggest solutions to open issues. They can also expand on suggestions of other attendees which they would not have considered otherwise. At the end of the meeting, stakeholders should agree on a concrete vision of the system. The importance of discussions with other stakeholders is underlined by the acceptance of \textbf{H1$_1$}.

Furthermore, synchronous viewings provide a clear time frame for requirements engineers.
Planned meetings provide clear start and end times. A well structured meeting should lead to satisfying results within the set time frame. This makes it easier for requirements engineers to foresee when they can expect the results of the vision video viewings.

\paragraph{Disadvantages}
One of the major difficulties when organizing synchronous meetings with many attendees is finding a time that is suitable for everyone. Vision videos can be watched with only a subset of all relevant stakeholders. However, this introduces the risk of missing out on ideas, especially ideas that build upon suggestions of others.

Moreover, conjoined meetings often are of a fixed length. This leads to discussions having to be cut short in order to maintain the meeting's schedule.

Another disadvantage is the fact that synchronous viewings also bear the risk of not gathering all possible arguments from shy stakeholders. In meetings with many attendees, some participants are likely to be more active than others. Other stakeholders may be reluctant to express new ideas or disagree with their peers. Stakeholders can resign to just repeating what other attendees mentioned instead of thinking about new ideas themselves.

%Another, albeit minor, disadvantage is the fact that  Synchronous meetings allow for attendees to sit back and simply agree with ideas mentioned by more extroverted peers.

%The use of webcams can also help simulate traditional meetings more directly.

\subsection{Pros and Cons of Asynchronous Viewing (RQ2)}
As for asynchronous viewings, we tested our method with ten test subjects. We found the following advantages and disadvantages:

%Asynchronous Watching
\paragraph{Advantages}
One of the main advantages of asynchronous viewings of vision videos is the fact that stakeholders can freely choose a time at which they watch the video. Requirements engineers do not need to find a single time that is suitable for all stakeholders. However, it is advisable to set a larger time frame in which the video should be watched.

Another benefit is that stakeholders have full control over the video during these asynchronous viewings. They can move freely along the video’s time line and explore its content at their own pace. This also allows them to re-watch individual sequences or even the whole video. Stakeholders can record their own ideas and concerns without inhibition. Our acceptance of \textbf{H2$_1$} emphasizes this advantage.

Asynchronous viewings of vision videos are a way to detect conflicts within the visions of the video’s creators and stakeholders. The method requires minimal effort from requirements engineers as they do not need to be present while the video is watched. Conflicts can then be addressed in meetings that should be shorter than completely synchronous viewings. Providing stakeholders with the video before a meeting also means that they have more time to prepare thoughtful suggestions. This could benefit the requirements engineering process.

\paragraph{Disadvantages}
The most important drawback of asynchronous viewing is that stakeholders miss out on the possibility to directly discuss suggestions and concerns with their peers. They can not respond to issues raised by others or expand on their ideas.

The missing discussion also means that conflicts in the visions of stakeholders can not be resolved immediately. While a single stakeholder can raise their concerns with the video’s creators and start a discussion with them, they are unable to confer with other stakeholders. This means that conflicting mental models between such stakeholders would need to be resolved in a separate manner.

Another disadvantage of asynchronous viewing is the fact that stakeholders need to record their thoughts themselves. This means that a way to record stakeholders’ comments is needed. Such a collaborative system could for example work through annotations that viewers can add to relevant video sections.

Additionally, stakeholders might not understand the video’s contents correctly. Such misunderstandings can not be rectified immediately when the video is watched asynchronously. Instead, another discussion, at least via e-mail, is required.

\subsection{Research Goal}
%beide haben vor und nachteile
%beste Nutzung (wenn möglich)
%crowdRE: asynchronous mit ticket system oder so
Both methods of watching vision videos presented in this paper have major advantages and disadvantages. It is not possible to objectively decide which method is superior.

Ultimately, the two presented methods are best suited for different use cases. Synchronous viewings most closely simulate the traditional use of vision videos. They fulfill the same purpose and should develop similar results. Asynchronous viewings can support the requirements elicitation process when it is hard to find a time at which all stakeholders can attend the same meeting.

A combination of the two methods could present the most suitable method for using vision videos in online settings. By sharing the video with stakeholders they can prepare themselves for the conjoined meeting. Misalignments between the vision presented in the video and a stakeholder's mental model can be detected before the meeting, meaning that they can point to the relevant video sections. Additionally, they can also develop ideas expanding on the video's contents or decide on any shown variants. This way, the synchronous meeting can be reduced to a compact discussion between stakeholders which in turn should lead to more ideas being discussed in a shorter amount of time.

The most important use case for asynchronous meetings is the elicitation of requirements from crowds. When dealing with large numbers of stakeholders, it is often impossible to organize and conduct a synchronous meeting. Discussions between individual stakeholders are also impossible due to the scale of the requirements engineering process. Letting crowds of stakeholders watch vision videos asynchronously enables each individual stakeholder to contribute their ideas and suggestions. A system recording these comments is still necessary but can also include automatic aggregations of the given data. This way, requirements engineers can easily gather high quality data from a crowd of stakeholders.

%Limitierte Zeit bei Gruppe
%Stakeholder-Rolle unbekannt
%niemand kannte das video
%wo "Probanden kannten sich nicht"-Teil?
%Unverständnis einiger Varianten problematisch -> haben vor abstimmung noch erklärt
\subsection{Threats to Validity}
There are a number of limitations to our results. In this section, we discuss these threats to validity according to the four types defined by Wohlin et al.~\cite{wohlin_experimentation_2012}.

\paragraph{Construct Validity}
%Construct validity is the extent to which the measurements used, often questionnaires, actually test the hypothesis or theory they are measuring. --> Brauche Ergebnisse, welche Ergebnisse nennen wir überhaupt?
The construct validity of our results is threatened by the fact that the answers to our questionnaire heavily rely on self reporting. Aspects like the tendency to change one's opinion after discussions with peers were not measured by any objective metrics. Instead, participants simply reported on a likert scale.
Another threat is that during individual sessions, participants may not understand the video's content, so that they cannot make immediate decisions after watching (see figure~\ref{fig:Experiment Design}). To mitigate this threat, the researcher asked the participant to recall the variants first. If they still have difficulty making a decision, the interviewer provided necessary descriptions in key words. 
Moreover, bad video quality can lead to contents not being shown clearly. This may lead to requirements being misunderstood. To check whether participants perceived video quality to be an issue, we asked them question about video quality. As results in section~\ref{addtional-result} show, most participants rated the video quality positively. This indicates that this threat did not exist in the context of our experiment.

\paragraph{Internal Validity}
% Fatigue or boredom of the participants may have affected the results of our experiment - to minimize this possible negative aspect we have kept the execution time as short as possible - maximum of 30 minutes
% outcome is not objective - opinion of the participants
% Repeating the experiment --> maybe a different outcome
The results of our experiment could have been effected by participants' fatigue or boredom. To minimize this threat to validity we kept the execution time of our experiment as short as possible. Not a single participant took longer than 30 minutes. 
Another threat to the internal validity of our results is that the outcome of discussions between participants is not objective. Arguments given for or against a particular variant shown in the vision video are solely based on the opinion of the participants. Repeating the experiment with a different set of test subjects might lead to a different outcome. However, we do not expect a change to the observed advantages and disadvantages of each method.

\paragraph{Conclusion Validity}
% small sample size --> maybe false conclusions
%Argumente bei asynchronem Schauen nicht distinkt!
A major threat to the conclusion validity of our results is the small sample size. Our preliminary experiment only included 16 participants. We still found a number of advantages and disadvantages of both methods presented in this paper.
Moreover, for the results of participants taking part in the individual interviews, we counted all arguments mentioned. We did not differentiate between already mentioned points and completely new notions. Nevertheless, arguments that are expressed multiple times still present meaningful feedback. Concerns shared by a number of stakeholders can be seen as more important than those only mentioned once.
 
\paragraph{External Validity}
%The presented study is only a preliminary one. Our results originate from a limited number of participants. 
Our selection of participants also threatens the external validity of this research. Most of them study or work in STEM related fields. As for the attendees of our group session simulating synchronous viewing of vision videos, we purposefully selected participants who did not know each other before the experiment. This ensured that participants did not have an inclination to respond to particular other participants who they had already discussed other topics with.
%viele haben schonmal an softwareentwicklungsprozessen teilgenommen
Furthermore, most participants of our experiment revealed that they had been involved in software development processes before taking part. They are more likely to be familiar with the role of stakeholders. Conducting the experiment with participants less familiar with development processes might lead to different discussions. However, we do not expect this to have impacted the usefulness of our methods.

\section{Conclusion}
\label{sec:concl}
Vision videos can be used as a way to create a shared understanding between stakeholders. The video itself can be used as the foundation for discussions on the design of a new system. However, the use of vision videos in times of social distancing has not been fully explored in scientific research.

In this paper, we present concepts for two different methods of using vision videos in online settings. We expand on the advantages and disadvantages of both synchronous and asynchronous viewing of videos and conduct a preliminary experiment. We observed participants watching a vision video in a between-groups design. One group watched the video synchronously, while the other group watched in individual sessions.

Our results indicate clear advantages of both methods. Synchronous viewings most closely simulate traditional vision video meetings where all relevant stakeholders meet in person to watch the video. Asynchronous viewings enable vision videos to be used with large numbers of stakeholders, when the sheer number of participants would prohibit a conjoined meeting due to scheduling constraints. Using vision videos and asynchronous viewing, requirements can be elicited from crowds. Both methods facilitate distributed teams to use vision videos without having to meet in person and therefore without major travel costs.
%However, we found that asynchronous viewings alone do not fulfill the same purpose as traditional vision video meetings. A synchronous, albeit shorter, meeting is necessary to solve conflicts in stakeholder's mental models. Nevertheless, asynchronous viewings are suited for use cases in which synchronous meetings are impossible, for example due to the large number of stakeholders.

The experiment presented in this paper is only a preliminary one. A larger experiment with practitioners would lead to more refined results. We also seek to evaluate whether our concepts are transferable to in-person uses of vision videos. Watching the video asynchronously before a shorter synchronous meeting could reduce the time required of stakeholders. %Additionally, ideas that might not be mentioned in a large meeting due to shyness or a scarcity of time could surface if our concepts are applied in offline settings. 
The use of our concepts for crowds of stakeholders should also be investigated further. Finally, a collaborative system to collect the feedback from stakeholders who watch the vision video asynchronously is yet to be developed.

\section*{Acknowledgment}

This work was supported by the Deutsche Forschungsgemeinschaft (DFG) under Grant No.: 289386339, project ViViUse.

\bibliography{Literature.bib}{}
\bibliographystyle{IEEEtran}

\end{document}